\documentclass[11pt,a4paper]{article}
\usepackage{jheppub}
\usepackage[T1]{fontenc}
\usepackage{amssymb}
\usepackage{mathtools}
\usepackage{stackengine}
\usepackage{scalerel}

\usepackage{hyperref}
\usepackage{epsfig}
\usepackage{amsmath,latexsym,amssymb}
\usepackage{graphicx}
\usepackage{braket}
\usepackage{pdfsync}
\usepackage{framed}
\usepackage{mathtools}

\usepackage{jheppub}
\usepackage[T1]{fontenc}
\usepackage{amssymb}
\usepackage{mathtools}
\usepackage{amsmath}
\usepackage{bm}
\usepackage{stackengine}

\usepackage{scalerel}
\usepackage{eufrak}
\usepackage{framed}

\usepackage{mathtools}

\usepackage{pstricks}

\usepackage{amsmath}
\usepackage{bbold}
\usepackage{bm}
\usepackage{latexsym}
\usepackage{braket}
\usepackage{slashed}
\usepackage{graphicx,booktabs,multirow}
\usepackage{eufrak}
\numberwithin{equation}{section}
\usepackage{color}
\usepackage{pstricks}
\usepackage{amssymb}

\makeatletter
\newcommand{\doublewidetilde}[1]{{%
  \mathpalette\double@widetilde{#1}%
}}
\newcommand{\double@widetilde}[2]{%
  \sbox\z@{$\m@th#1\widetilde{#2}$}%
  \ht\z@=.9\ht\z@
  \widetilde{\box\z@}%
}
\makeatother

\usepackage{hyperref}
\usepackage{slashed}
\usepackage{epsfig}
\usepackage{amsmath,latexsym,amssymb}
\usepackage{graphicx}
\usepackage[latin1]{inputenc}
\usepackage{braket}
\usepackage{pdfsync}
\usepackage{framed}

\usepackage{mathtools}

\def\be{\begin{equation}}
\def\ee{\end{equation}}
\def\ba{\begin{eqnarray}}
\def\ea{\end{eqnarray}}

\newcommand{\bz}{\bar{z}}
\newcommand{\bw}{\bar{w}}
\newcommand{\bh}{\bar{h}}

\newcommand{\by}{\bar{y}}
\newcommand{\bx}{\bar{x}}

\newcommand{\zbar}{\bar{z}}

\newcommand{\hy}{{}_{2}F_1}

\def\d{\delta}
\def\D{\Delta}

\def\cM{{\cal M}}

\newcommand{\comment}[1]{}

\newcommand{\eea}{\end{eqnarray}}

\author{
Wei Fan${}^{1}$, Angelos Fotopoulos${}^{2,3}$, Stephan Stieberger${}^{4}$,
Tomasz R.\ Taylor${}^{2}$,\, Bin Zhu${}^2$\\[0.5cm] $^1$\it
Department of Physics, School of Science, Jiangsu University of Science and Technology,
Zhenjiang, 212003, China\\[0.2cm]
 $^2${\it Department of Physics \\
  Northeastern University, Boston, MA 02115, USA}\\[0.2cm]
 $^3${\it Department of Sciences\\
Wentworth Institute of Technology, Boston, MA 02115, USA} \\[.2 cm]
$^4${\it Max--Planck--Institut f\"{u}r Physik,	Werner--Heisenberg--Institut, \\80805 M\"unchen, Germany}}

\title{\boldmath Conformal Blocks from Celestial Gluon Amplitudes  \unboldmath}

\abstract{In celestial conformal field theory, gluons are represented by primary fields with dimensions $\Delta=1+i\lambda$, $\lambda\in\mathbb{R}$ and  spin $J=\pm 1$, in the adjoint representation of the gauge group. All two- and three-point correlation functions of these fields are zero as a consequence of four-dimensional kinematic constraints. Four-point correlation functions contain delta-function singularities enforcing planarity of four-particle  scattering events. We relax these constraints by taking a shadow transform of one field  and perform  conformal block decomposition of the corresponding correlators. We compute the conformal block coefficients. When decomposed in channels that are ``compatible'' in two and four dimensions, such four-point correlators contain conformal blocks of primary fields with dimensions $\Delta=2+M+i\lambda$, where $M\ge 0$ is an integer, with integer spin $J=-M,-M+2,\dots,M-2,M$. They appear in all gauge group representations obtained from a tensor product of two adjoint representations.   When decomposed in incompatible channels, they also contain primary fields with continuous complex spin, but with positive integer dimensions.}


\begin{document}
\maketitle
\section{Introduction}
Celestial conformal field theory (CCFT) links the amplitude program with flat holo\-graphy. It grew from the observation that the symmetries of asymptotically flat spacetime \cite{Bondi:1962px,Sachs:1962wk,Barnich:2009se}, which act on incoming and outgoing particles, are also present at the level of the scattering S-matrix \cite{Strominger:2013jfa}. The ultimate goal of CCFT is to describe the four-dimensional world as a hologram on a celestial sphere \cite{Strominger:2017zoo}, in the framework of two-dimensional conformal field theory (CFT). To that end, the scattering amplitudes are recast into the form of CFT correlation functions \cite{Strominger1706} by changing the basis of asymptotic states from the standard momentum basis to the basis of conformal wave packets, a.k.a.\ the boost basis \cite{Pasterski1705}.

After the  amplitudes are converted to the boost basis, they transform under $SL(2,C)$ Lorentz transformations as two-dimensional CFT correlators of primary fields. Four-dimen\-sional soft theorems \cite{Weinb,Cachazo:2014fwa}
acquire a form of Ward identities of holomorphic and antiholomorphic currents
\cite{Strominger1401,Strominger1609,Strominger1810,Fan1903,Strominger1904, Mason1905,Puhm1905,Guevara1906}. The subleading soft graviton theorem is of particular importance because it allows identifying two-dimensional energy-momentum tensor and Virasoro operators, which generate conformal transformations, a.k.a.\ superrotations of primary fields on celestial sphere \cite{Foto1906}. The leading graviton theorem is related to supertranslations which together with superrotations generate the algebra of BMS symmetry \cite{Foto1912}. Another important component are collinear theorems \cite{tt} which allow studying the limits of operator insertions at coinciding points, hence deriving the operator product expansions of the operators associated to gluons and gravitons \cite{Fan1903,Strominger1910,Ebert:2020nqf,Banerjee:2020kaa,Foto1912}.

Further progress in CCFT depends on understanding the spectrum of primary fields and their interactions. In standard CFT \cite{DiF} this is handled by using conformal block decomposition of the correlation functions. The importance of CCFT block decomposition has been recognized in Refs.\cite{Lam:2017ofc,Volovich1904,Law:2020xcf} and approached there from various angles. All this work had to face the same obstacle: since CCFT correlators originate from scattering amplitudes, they are highly constrained by four-dimensional kinematics. For example, all three-gluon amplitudes vanish on-shell while four-gluon amplitudes contain a singular term enforcing planarity of four-particle scattering events.

We will approach this problem from a different angle. In our recent studies, the shadow transform \cite{Osborn:2012vt} has played a prominent role in establishing the connection between soft theorems and the Ward identities of symmetries in CCFT \cite{Foto1912,Fotopoulos:2020bqj}. We will use it here to relax the kinematic constraints on four-gluon amplitudes. After one gluon is replaced by the shadow field, the singularities disappear and the amplitude acquires a simple form that can be analyzed by using standard CFT techniques.

The paper is organized as follows. In Section 2, we discuss CCFT in the context of standard radial quantization and point out some peculiar features that make it so different from a ``garden variety'' of CFTs. In Section 3, we apply shadow transform to one of the gluon fields and set the groundwork for conformal block decomposition. We extract the blocks in Section 4. In general case, this is quite a tedious process, however it becomes a simple task in the limit of soft shadow field. In this limit, the antiholomorphic part degenerates to a single antichiral weight. We conclude in Section 5.

\section{Radial quantization of CCFT}
In radial quantization of two-dimensional conformal field theory (CFT) \cite{DiF}, the asymptotic states characterized by dimensions $\Delta$ and spin $J$ are created by acting on the vacuum state with primary quantum field operators:
\be
|\Delta,J;{ in}\rangle=\lim_{z,\zbar\to 0}\phi_{\Delta,J}(z,\zbar)|0\rangle\ ,\qquad{}\langle\Delta,J;out|=\lim_{z,\zbar\to 0}\langle 0|\phi_{\Delta,J}^\dagger(z,\bz)\ .
\label{inout}\ee
In terms of the chiral weights $h$ and $\bh$, $\Delta=h+\bh$
and $J=h-\bh$. When $\Delta$ and $J$  are real, the hermitian conjugate field is defined by
\be[\phi_{\Delta,J}(z,\zbar)]^\dagger=\bz^{-2h}z^{-2\bh}\phi_{\Delta,J}(1/\zbar,1/z)\ .\label{hc}\ee
Following this definition, the inner product is
\be\langle{\Delta, J};in|{\Delta,J};out\rangle=\lim_{\xi,\bar\xi\to\infty}\bar\xi^{2h}\xi^{2\bh}\langle 0|\phi_{\Delta,J}(\bar\xi,\xi)\phi_{\Delta,J}(0,0)|0\rangle\ .\ee
With the standard form of conformally covariant two-point function given by
\be\langle 0|\phi_{\Delta,J}(z,\bz)\phi_{\Delta,J}(w,\bw)|0\rangle=     \frac{C}{(z-w)^{2h}(\bz-\bw)^{2\bh}},\label{tpt}\ee
this inner product becomes
\be\langle{\Delta, J};in|{\Delta,J};out\rangle=C\ .\ee

In celestial CFT (CCFT) with $\Delta=1+i\mathbb{R}$, one can consider an alternative definition of hermitean conjugation -- by relating fields with complex conjugate dimensions belonging to the principal continuous series. It can be done by using shadow transforms.
The shadow of a primary field with chiral weights $h,\bh$, hence with the dimension $\Delta=h+\bh$ and spin
$J=h-\bh$, is defined \cite{Osborn:2012vt} as
\be\tilde\phi(z,\bz)=k_{h,\bh}\int d^2y(z-y)^{2h-2}(\bz-\by)^{2\bh-2}\phi(y,\by)\ ,~ k_{h,\bh}=(-1)^{2(h-\bh)}{\Gamma(2-2h)\over\pi\Gamma(2\bh-1)}\label{shdef} .\ee
The shadow field $\tilde\phi(z,\bz)$ is also a primary, but with weights $1-h,1-\bh$, hence with the dimension $\tilde\Delta=2-\Delta$ and spin $ \tilde J=-J$. The constant $k_{h,\bh}$ is chosen in such a way that, for integer or half-integer spin, $\tilde{\tilde\phi}(z,\bz)=\phi(z,\bz)$.  Note that for Re$\Delta=1$,
$\tilde\Delta=\Delta^*$. In this case, we can define
\be[\phi_{\Delta,J}(z,\zbar)]^\dagger=\bz^{-2h}z^{-2\bh}\widetilde{\phi_{\Delta^*,-J}}(1/\zbar,1/z)
=\bz^{-2h}z^{-2\bh}\tilde{\phi}_{\Delta,J}(1/\zbar,1/z)~~~~(\Delta=1+i\mathbb{R})
 .\label{hc1}\ee
The two-point function describing a primary field propagation into the shadow is constrained by conformal covariance to
\be\langle 0|\tilde{\phi}_{\Delta,J}( z,\bar{z})\phi_{\Delta,J}(w,\bw)|0\rangle=     \frac{\widetilde C}{( z-w)^{2h}(\bar{z}-\bw)^{2\bh}}\ ,\label{tpsh}\ee
leading to
\be\langle{\Delta, J};in|{\Delta,J};out\rangle=\widetilde C\ .\ee
Note that Eq.(\ref{tpsh}) is equivalent to
\be\langle 0|\phi_{\Delta^*,-J}(z,\bz)\phi_{\Delta,J}(w,\bw)|0\rangle=\widetilde{C}k^{-1}_{1-h,1-\bh}\, \delta^{(2)}(z-w)~~~~(\Delta=1+i\mathbb{R})\ .\label{tpts}\ee

CCFT correlators originate from four-dimensional scattering amplitudes with the asymptotic states transformed to the basis of conformal wave packets \cite{Pasterski1705}. Massless gluons are associated to primary fields $\phi^a$ labeled by indices $a$ transforming in the adjoint representation of the gauge group. Three-gluon amplitudes vanish as a consequence of kinematic constraints \cite{tt} therefore the respective three-point CCFT correlators are zero. On the other hand, as shown in Refs.\cite{Fan1903,Strominger1910}, the operator product expansions (OPE) of $\phi^a(z)\phi^b(0)$ contain gluons: $\phi^a(z)\phi^b(0)\sim z^{-1}f^{abc}\phi^c(0)+\dots$ These two results are compatible if and only if $C=0$,\footnote{This can be shown directly at the level of celestial correlators by using constraints of Poincar\'e invariance \cite{Law:2019glh}.} which means that the primary fields associated to gluons do not propagate into each other.\footnote{This argument does not exclude though $\widetilde C\neq 0$ because a contact term like (\ref{tpts}) may be not reachable as a limit of OPE.} This is not necessarily a disaster, provided that we can identify ``missing'' fields. A good place to start are the four-point correlators related to four-gluon celestial amplitudes. Unfortunately here again, we are to a rough start because
such amplitudes have their support restricted by momentum conservation. They contain delta-function contact terms enforcing planarity of four-gluon scattering events \cite{Strominger1706}. We will relax kinematic constraints by choosing one of the fours fields to be a shadow field. Such a shadow transformation leads to standard CFT correlators,
well-defined over the whole domain of points on ${\cal CS}_2$. These can be analyzed by using standard CFT tools including the conformal block decomposition. Another reason for studying ``shadowed'' celestial amplitudes is the connection between adjoint operators and shadow fields, Eq.(\ref{hc1}), which could relate four-dimensional (asymptotic) shadow wave functions to {\it in\/} and {\it out\/} states (\ref{inout}) of two-dimensional CFT.\footnote{Similar ideas have been considered in \cite{spas}.}
\section{Four points with one shadow}
\subsection{CCFT correlators from celestial amplitudes}
Celestial amplitudes are obtained from conventional scattering amplitudes by performing Mellin transforms with respect to light-cone energies.
The connection between light-like four-momenta $p^\mu$ of massless particles and points $z\in {\cal C S}^2$ relies on the following  parametrization:
\be\label{momparam}
p^\mu= \omega q^\mu, \qquad q^\mu={1\over 2} (1+|z|^2, z+\bz, -i(z-\bz), 1-|z|^2)\ ,
\ee
where $\omega$ is the light-cone energy ($\omega=E+p_z,~p_x+ip_y=\omega z$), and $q^\mu$ are null vectors -- the directions along which the massless state propagates. In Lorentzian spacetime endowed with (1,3) metric,  complex variables $z$ (with $\bz=z^*$) parameterize Euclidean ${\cal C S}^2$.

In CCFT, celestial amplitudes are identified with the correlators of primary fields. Here, we use the following identification:
\begin{eqnarray}\label{map}
\Big\langle \prod_{n=1}^N&&\!\!\!\!\!\!\phi_{\D_n,J_n}^{a_n}(z_n,\zbar_n)\Big\rangle =\\ \nonumber &=& \sum_{\epsilon_2,\dots,\epsilon_{N}=\pm 1}\! \big({\prod_{n=1}^N}   \int_0^\infty\!\!\! d \omega_n  \omega_n^{\D_n-1} \big) \d^{(4)}\big(\omega_1q_1+\!\sum_{i=2}^{N}\epsilon_i\omega_iq_i \big)\cM(\omega_n, z_n, \bz_n, J_n, a_n),
\end{eqnarray}
where $\cM(\dots, J_n,a_n)$ are Feynman's matrix elements for particles with helicities $J_n$ identified with the spin of operators and indices $a_n$ in the adjoint representation of the gauge group. The r.h.s.\ of Eq.(\ref{map}) involves the sum over parameters $\epsilon$ which take values ${+}1$ and ${-}1$ for outgoing and incoming particles, respectively. In other words, we are summing over all scattering channels and in order to avoid double counting, we assume that particle number 1 is outgoing. The helicities $J_n$ (and colors $a_n$) are counted however as if all particles were outgoing -- they take the same values in all channels. It should be stressed that there is no information lost when summing over channels because each of them covers a different patch of celestial coordinates. The mapping (\ref{map}) covers the entire range of celestial coordinates allowed by momentum conservation and encodes all channels in a single CCFT correlator. If one wants to consider each channel separately, one needs to make a distinction between two-dimensional primary fields representing incoming and outgoing particles in four dimensions. This would obviously complicate  crossing symmetries and conformal bootstrap, therefore we prefer remaining uncommitted to making such a distinction until we obtain the correlators and proceed with the conformal block decomposition.
\subsection{Shadow transform of the four-gluon correlator}
The four-point CCFT correlator corresponding to four-gluon celestial amplitude in ``mostly plus'' MHV helicity configuration is given (up to a numerical factor) by \cite{Strominger1706}\footnote{We are using the color basis of Ref.\cite{DelDuca:1999rs}.}
\begin{align}\nonumber
\Big\langle\phi_{\D_1,-}^{a_1}&(z_1,\zbar_1)\phi_{\D_2,-}^{a_2}(z_2,\zbar_2)\phi_{\D_3,+}^{a_3}(z_3,\zbar_3)
\phi_{\D_4,+}^{a_4}(z_4,\zbar_4)\Big\rangle=~\delta(\sum_{i=1}^4\lambda_i)\prod_{i<j}(z_{ij}\bar z_{ij})^{-i\frac{\lambda_i}{2}- i\frac{\lambda_j}{2}}\\ &\times\!\delta(z-\bar z)
\Big(\frac{z_{12}}{z_{13}z_{24}z_{34}}\Big)\Big(\frac{\bz_{34}^2}{\bz_{13}\bz_{24}\bz_{14}\bz_{23}}\Big)
 (f^{a_1a_2b}f^{a_3a_4b}-zf^{a_1a_3b}f^{a_2a_4b})\ ,
\label{glu4}
\end{align}
with the cross-ratios
\begin{equation}
z=  \frac{z_{12}z_{34}}{z_{13}z_{24}}\ ,\ \bar z=  \frac{\bar z_{12}\bar z_{34}}{\bar z_{13}\bar z_{24}}\ .
\label{crossr}\end{equation}
The delta function $\delta(z-\bar z)$ enforces planarity of four-particle scattering while the other one, $\delta(\sum_{i=1}^4\lambda_i)$, appears as a consequence four-dimensional conformal invariance which holds in Yang-Mills theory at the tree level only \cite{Stieberger1812}. The latter one gives rise to an infinite factor $\delta(0)$ when the correlator is evaluated at $\sum_{i=1}^4\lambda_i=0$ and will be ignored in the following discussion.
Recall that the chiral weights of gluon fields are:
\be
\begin{array}{cc} (h_1,\bh_1)=(\frac{i\lambda_1}{2},1+\frac{i\lambda_1}{2}) \ ,\qquad& (h_2,\bh_2)=(\frac{i\lambda_2}{2},1+\frac{i\lambda_2}{2})\ ,\\[3mm]
(h_3,\bh_3)=(1+\frac{i\lambda_3}{2},\frac{i\lambda_3}{2}) \ ,\qquad& (h_4,\bh_4)=(1+\frac{i\lambda_4}{2},\frac{i\lambda_4}{2})\ .
\end{array}
\ee
All these fields have integer spin, therefore that correlator (\ref{glu4}) is a {\it single-valued\/} function of four complex variables $z_i$. This implies, in particular, that there are {\it no\/} monodromy factors appearing when crossing scattering channels.

We are interested in the shadow correlator
\begin{align}
\Big\langle&\tilde\phi_{\tilde\D_1,+}^{a_1}(z_{1'},\zbar_{1'})\phi_{\D_2,-}(z_2,\zbar_2)\phi_{\D_3,+}(z_3,\zbar_3)
\phi_{\D_4,+}(z_4,\zbar_4)\Big\rangle~=\label{shadef}\\
& \int \frac{d^2z_1}{(z_1-z_1')^{2-i\lambda_1}(\bar z_1-\bar z_1')^{-i\lambda_1}}
\Big\langle\phi_{\D_1,-}^{a_1}(z_1,\zbar_1)\phi_{\D_2,-}^{a_2}(z_2,\zbar_2)\phi_{\D_3,+}^{a_3}(z_3,\zbar_3)
\phi_{\D_4,+}^{a_4}(z_4,\zbar_4)\Big\rangle .\nonumber
\end{align}
We will express it in terms of the cross ratios
\begin{equation}
z'=  \frac{z_{1'2}z_{34}}{z_{1'3}z_{24}}\ ,\ \bar z'=  \frac{\bar z_{1'2}\bar z_{34}}{\bar z_{1'3}\bar z_{24}}\ .
\label{crossr1}\end{equation}
Note that the shadow field has chiral weights $(h_{1'},\bh_{1'})=(1-\frac{i\lambda_1}{2},-\frac{i\lambda_1}{2})$.
In order to perform the shadow integral (\ref{shadef}), we change the integration variables from $z_1$ to \begin{equation}y=\frac{z}{z'}= \frac{z_{12}z_{1'3}}{z_{1'2}z_{13}}.\label{yvar}\ee The corresponding Jacobian is $|dz_1/dy|^2$ with
\be \frac{d z_1}{d y} = \kappa^2 z_{23}z_{1'2}z_{1'3}\ ,\qquad \kappa =\frac{z_{13}}{z_{1'3}z_{23}}=( y\, z_{21'} +z_{1'3})^{-1} \ee
In terms of the new variable: \begin{align}\nonumber
z_{12} = \kappa \, y \, z_{23} z_{1'2},\quad &\quad
z_{13} = \kappa \, z_{23} z_{1'2} ,\\ z_{14} = \kappa\, (y-\frac{1}{z'})\, z_{34} z_{1'2},\quad &\quad
z_{11'} = \kappa\, (y-1) \,z_{1'2} z_{1'3}.
\end{align}
In this way, we obtain
\begin{align}\nonumber
\Big\langle\tilde\phi_{\tilde\D_1,+}^{a_1}&(z_{1'},\zbar_{1'})\phi_{\D_2,-}^{a_2}(z_2,\zbar_2)\phi_{\D_3,+}^{a_3}
(z_3,\zbar_3)
\phi_{\D_4,+}^{a_4}(z_4,\zbar_4)\Big\rangle\\[1mm] &= ( z_{1'2}\bar{z}_{1'2})^{\frac{i\lambda_1}{2}-\frac{i\lambda_2}{2}} (z_{1'3}\bar{z}_{1'3} z_{34}\bar{z}_{34} )^{-\frac{i\lambda_3}{2}-\frac{i\lambda_4}{2}} (z_{23}\bar{z}_{23})^{i\lambda_4} (z_{24}\bar{z}_{24})^{\frac{i\lambda_3}{2}-\frac{i\lambda_4}{2}} \label{glu5}  \\[1mm] \nonumber  &~~~~~~\times\!
\Big(\frac{z_{23}}{z_{1'3}^2z_{24}z_{34}}\Big)\Big(\frac{\bz_{1'2}\,\bz_{34}^2}{\bz_{1'3}\bz_{23}\bz_{24}^2}\Big)
 \,\Big[f^{a_1a_2b}f^{a_3a_4b}I(z')+f^{a_1a_3b}f^{a_2a_4b}\tilde I(z')\Big]\ ,
\end{align}
where
\begin{align}\label{ibegin}
I(z')=&\int d^2y\, \delta(z'y-\bar{z}'\bar{y}) \,y \,(y \bar{y})^{-\frac{i\lambda_1}{2}-\frac{i\lambda_2}{2}}(y-1)^{-2} [(y-1)(\bar{y}-1)]^{i\lambda_1}\\[-1mm]&~~~~~~~~\times\! (\bar{z}' \bar{y}-1)^{-1}[(\bar z'\bar y-1)(z'y-1)]^{-\frac{i\lambda_1}{2}-\frac{i\lambda_4}{2}}\ ,\nonumber \\[1mm]
\tilde I(z')=&-z'\!\!\int d^2y\, \delta(z'y-\bar{z}'\bar{y}) \,y^2 \,(y \bar{y})^{-\frac{i\lambda_1}{2}-\frac{i\lambda_2}{2}}(y-1)^{-2} [(y-1)(\bar{y}-1)]^{i\lambda_1}\\[-1mm]&~~~~~~~~~~~\times\! (\bar{z}' \bar{y}-1)^{-1}[(\bar z'\bar y-1)(z'y-1)]^{-\frac{i\lambda_1}{2}-\frac{i\lambda_4}{2}} \ .\nonumber
\end{align}
We parametrize $y$ as $y=r e^{i\phi}$ and integrate over the complex plane in radial coordinates. We also parametrize $z'= R e^{i\theta}$, so that $R=\sqrt{z'\bz'}$ and $e^{i\theta}=\sqrt{z'/\bz'}$.  From the angular integration, $\delta(z'y-\bar{z}'\bar{y})$ picks two values: $\phi = -\theta$ and $\phi = -\theta+\pi$ $(e^{i\phi}=\pm e^{-i\theta})$. In this way, the integrals split as $I=I_++I_-$ and $\tilde I=\tilde I_++\tilde I_-$, with
\begin{equation}
I_{+} = \frac{1}{z'} \int_0^{+\infty} dr \,r^{1-i\lambda_1-i\lambda_2} \Big [ r \Big(\frac{\bar{z}'}{z'}\Big)^{\frac{1}{2}}-1\Big]^{-2+i\lambda_1} \Big[ r \Big( \frac{z'}{\bar{z}'}\Big)^{\frac{1}{2}} -1 \Big]^{i\lambda_1} \, ( r\sqrt{z'\bar{z}'}-1)^{-1-i\lambda_1-i\lambda_4} \,\, ,\label{eq:Ir1}
\end{equation}
\begin{equation}
I_{-} = \frac{1}{z'} \int_0^{+\infty} dr \,r^{1-i\lambda_1-i\lambda_2} \Big [ r \Big(\frac{\bar{z}'}{z'}\Big)^{\frac{1}{2}}+1\Big]^{-2+i\lambda_1} \Big[ r \Big( \frac{z'}{\bar{z}'}\Big)^{\frac{1}{2}} +1 \Big]^{i\lambda_1} \, ( r\sqrt{z'\bar{z}'}+1)^{-1-i\lambda_1-i\lambda_4} \,\, ,\label{eq:Ir2}
\end{equation}
and similar expressions for $\tilde I_\pm$.
It is convenient to change the integration variables to $r\sqrt{z'\bar{z}'}$. Then
\be I_{+} = \frac{(z'\bz')^{\frac{i\lambda_2}{2}}}{z'^2\bz'} \int_0^{+\infty} dr \,r^{1-i\lambda_1-i\lambda_2} \Big ( \frac{r}{z'}-1\Big)^{-2+i\lambda_1}\Big ( \frac{r}{\bz'}-1\Big)^{i\lambda_1} \, ( r-1)^{-1-i\lambda_1-i\lambda_4} \ ,\label{iplus}
\end{equation}
\begin{equation}
I_{-}  = \frac{(z'\bz')^{\frac{i\lambda_2}{2}}}{z'^2\bz'} \int_0^{+\infty} dr \,r^{1-i\lambda_1-i\lambda_2} \Big ( \frac{r}{z'}+1\Big)^{-2+i\lambda_1}\Big ( \frac{r}{\bz'}+1\Big)^{i\lambda_1} \, ( r+1)^{-1-i\lambda_1-i\lambda_4}\ .\label{iminus}
\end{equation}
By comparing this integration variable with (\ref{yvar}) we see that $r$ is the original cross-ratio $z$, {\it cf.\/} Eq.(\ref{crossr}). In $I_+$, the integration region splits into two four-dimensional scattering channels:
 $s\makebox{-channel}$ $(12\rightleftharpoons 34)_{\bm{\mathfrak{4}}}$ for $r\ge 1$ and
$ t\makebox{-channel}~ (13\rightleftharpoons 24)_{\bm{\mathfrak{4}}}$ with $r\in (0, 1)$.\footnote{Note that $s$, $t$ and $u$ channels refer here to the channels of celestial amplitudes with $s\ge$, $t \ge 0$ and $u\ge 0$, respectively, which are related by crossing symmetry. The term ``channel'' is sometimes used in a different way, to distinguish between the contributions of virtual particles propagating in these channels to one particular amplitude.} As mentioned before, there is no monodromy when crossing the boundary at $r=1$. On the other hand, $I_-$ comes from the $u\makebox{-channel}$ $(14\rightleftharpoons 23)_{\bm{\mathfrak{4}}}$. Hence
\be I(z')=I_s+I_t+I_u\ ,~~\qquad \tilde I(z')=\tilde I_s+\tilde I_t+\tilde I_u\ ,\label{iend}\ee
where subscripts indicate the contributions of respective scattering channels.

\section{Conformal blocks}
We first proceed with the conformal decomposition of the four-point correlator (\ref{glu5}) in the two-dimensional $(12\rightleftharpoons 34)_{\bm{\mathfrak{2}}}$ channel. To that end, we
set $z'=x$, $\zbar'=\bar x$ and define \cite{DiF}
\be
G_{34}^{21}(x,\bar{x}) =  \lim_{z_{1'}, \bar{z}_{1'}\rightarrow \infty} z_{1'}^{\,2 h_{1'}} \bar{z}_{1'}^{\,2 \bar{h}_{1'}}
\Big\langle\tilde\phi_{\tilde\D_1,+}^{a_1}(z_{1'},\zbar_{1'})\phi_{\D_2,-}^{a_2}(1,1)\phi_{\D_3,+}^{a_3}
(z'=x,\zbar'=\bar x)
\phi_{\D_4,+}^{a_4}(0,0)\Big\rangle\ .
\label{eq:G2134}
\ee
We obtain
\begin{align} G_{34}^{21}(x,\bx) ~=~ (1-x)^{1+i\lambda_4}& \, x^{-1-\frac{i\lambda_3}{2}-\frac{i\lambda_4}{2}} \,(1-\bx)^{-1+i\lambda_4} \, \bx^{\,2-\frac{i\lambda_3}{2}-\frac{i\lambda_4}{2}}\label{gg12} \\[1mm]\!\!\! &\!\!\!\!\!\times\!
\Big[f^{a_1a_2b}f^{a_3a_4b}I(x)+f^{a_1a_3b}f^{a_2a_4b}\tilde I(x)\Big]\ .\nonumber
 \end{align}
 As we will see later, the integrals $I$ and $\tilde I$ can be expressed in terms of the Appell function $F_1$. To avoid getting lost in details, we prefer to postpone this step and first consider  the soft limit of $\lambda_1=0$.
 \subsection{The shadow current}
{}For $\lambda_1=0~(\Delta_1=\tilde\Delta_1=1)$, we expect some simplifications to occur because one of the primary fields becomes the holomorphic shadow of the current $\bar{\mathfrak{j}}^{ a}(\bz)$ generating global gauge group transformations:
\be \tilde\phi_{\tilde\D_1=1,+}^{a_1}(z_{1'},\zbar_{1'})=-2\pi\, \tilde{\mathfrak{j}}^{ a_1}(z_{1'}),\ee
where \begin{equation}
\tilde{\mathfrak{j}}^a(z) =-\frac{1}{2\pi} \int \frac{d^2 w}{(z-w)^2} \, \bar{\mathfrak{j}}^a(\bar{w}) \,\, .
\end{equation}
The antiholomorphic current satisfies the following Ward identity \cite{Fan1903}:
\begin{align}
\Big\langle\, \bar{\mathfrak{j}}^{a_1}(\bar{w})& \phi^{a_2}_{\Delta_2, J_2}(z_2,\bar{z}_2) \cdots \phi^{a_N}_{\Delta_N, J_N}(z_N,\bar{z}_N) \Big\rangle \nonumber\\
&=\sum_{i=2}^N \sum_b \frac{f^{a_1 a_i b}}{\bar{w}-\bar{z}_i} \Big\langle  \phi^{a_2}_{\Delta_2, J_2}(z_2,\bar{z}_2) \cdots \phi^b_{\Delta_i,J_i}(z_i,\bar{z}_i)\cdots \phi^{a_N}_{\Delta_N, J_N}(z_N,\bar{z}_N)\Big\rangle .\label{step1}\end{align}
The shadow transform can be performed by using
\begin{equation}
\int \frac{d^2 w}{(z-w)^2(\bar{w}-\bar{z}_i)}=-\frac{2\pi}{z-z_i}\,\label{step3}
\end{equation}
and yields
\begin{align}
\Big\langle\, \tilde{\mathfrak{j}}^{a_1}({z})& \phi^{a_2}_{\Delta_2, J_2}(z_2,\bar{z}_2) \cdots \phi^{a_N}_{\Delta_N, J_N}(z_N,\bar{z}_N) \Big\rangle \nonumber\\
&=\sum_{i=2}^N \sum_b \frac{f^{a_1 a_i b}}{z-z_i} \Big\langle  \phi^{a_2}_{\Delta_2, J_2}(z_2,\bar{z}_2) \cdots \phi^b_{\Delta_i,J_i}(z_i,\bar{z}_i)\cdots \phi^{a_N}_{\Delta_N, J_N}(z_N,\bar{z}_N)\Big\rangle \,\, .\label{step2}\end{align}
At this level, it seems that the shadow current $\tilde{\mathfrak{j}}^{a}({z})$ can be identified with the holomorphic current $\mathfrak{j}^{a_1}({z})$. For $N=4$, however the r.h.s.\ of  Eq.(\ref{step1}) and Eq.(\ref{step2}) are zero. Nevertheless, as shown below, the limit of
$\lambda_1=0$ yields an ``almost holomorphic'' correlator with a simple antiholomorphic part.\footnote{This means that the soft limit and shadow transformations do not commute in the case of $N=4$, which is probably caused by the kinematic constraints on the operator insertion points.}
 \subsection{$\lambda_1=0$ soft limit}
In the limit of $\lambda_1=0$, the integrals (\ref{ibegin})-(\ref{iend}) can be expressed in terms of hypergeometric functions $\hy$. Note that in this limit, $\lambda_2+\lambda_3+\lambda_4=0$. The integral $I(x)=I_s(x)+I_t(x)+I_u(x)$, split into the contributions of three channels, is given by
\begin{align} I_s(x) ~=~ &\frac{(x\bx)^{\frac{i\lambda_2}{2}}}{x^2\bx} \int_1^{+\infty}\!\! dr\frac{ r^{1-i\lambda_2} }{(r-1)^{1+i\lambda_4}}\,\Big ( \frac{r}{x}-1\Big)^{-2}\nonumber \\[1mm]&=
\frac{(x\bx)^{\frac{i\lambda_2}{2}}}{\bx} B(1-i\lambda_3,-i\lambda_4)\,_2F_1\left({2,1-i\lambda_3\atop 1+i\lambda_2};x\right)
 \ ,\label{is0}\\ I_t(x) ~=~ &-\frac{(x\bx)^{\frac{i\lambda_2}{2}}}{x^2\bx} \int_0^{1}\!\! dr\frac{ r^{1-i\lambda_2} }{(1-r)^{1+i\lambda_4}}\,\Big ( \frac{r}{x}-1\Big)^{-2}\nonumber \\[1mm]&=
-\frac{(x\bx)^{\frac{i\lambda_2}{2}}}{x^2\bx}
B(2-i\lambda_2,-i\lambda_4)\,_2F_1\left({2,2-i\lambda_2\atop 2+i\lambda_3};\frac{1}{x}\right)  \ ,\label{it0}\\[1mm] I_u(x) ~=~ &\frac{(x\bx)^{\frac{i\lambda_2}{2}}}{x^2\bx} \int_0^{\infty}\!\! dr\frac{ r^{1-i\lambda_2} }{(1+r)^{1+i\lambda_4}}\,\Big ( \frac{r}{x}+1\Big)^{-2}\nonumber \\[1mm]&=
\frac{(x\bx)^{\frac{i\lambda_2}{2}}}{x^2\bx}
B(2-i\lambda_2,1-i\lambda_3)\,_2F_1\left({2,2-i\lambda_2\atop 3+i\lambda_4};1-\frac{1}{x}\right)
 \ ,\label{iu0}\\[1mm]&=
\frac{(x\bx)^{\frac{i\lambda_2}{2}}}{\bx}
B(2-i\lambda_2,1-i\lambda_3)  \,_2F_1\left({2,1-i\lambda_3\atop 3+i\lambda_4};\,1-x\right)
.\nonumber
\end{align}
Similarly, $\tilde I(x)=\tilde I_s(x)+\tilde I_t(x)+\tilde I_u(x)$, with
\begin{align}\label{tis}
\tilde{I}_{s}(x) &=-\frac{(x\bx)^{\frac{i\lambda_2}{2}}}{\bx} \,B(-i\lambda_3,-i\lambda_4)\,_2F_1\left({2,-i\lambda_3\atop i\lambda_2};\,x\right)\\[.1mm]
\label{tit}\tilde{I}_{t}(x)&=
\frac{(x\bx)^{\frac{i\lambda_2}{2}}}{x^2\bx} \,B(3-i\lambda_2,-i\lambda_4)\,_2F_1\left({2,3-i\lambda_2\atop 3+i\lambda_3}; \,\frac{1}{x}\right) \\[.1mm]
\label{tiu}\tilde{I}_{u}(x) &=-\frac{(x\bx)^{\frac{i\lambda_2}{2}}}{\bx}
\, B(3-i\lambda_2,-i\lambda_3) \,_2F_1\left({2,-i\lambda_3\atop 3+i\lambda_4};1-x\right)
\end{align}

In the first step, we want to decompose into conformal blocks the $s$-channel contribution to the correlator (\ref{gg12}). After taking the $\lambda_1=0$ limit in (\ref{gg12}) and using Eqs.(\ref{is0}) and (\ref{tis}), we obtain
\be G_{34}^{21}(x,\bx)_{s,\lambda_1=0} ~=~ (1-\bx)^{-1+i\lambda_4}\bx^{1+i\lambda_2}
\Big[f^{a_1a_2b}f^{a_3a_4b}{S}_{34}^{21}(x)+f^{a_1a_3b}f^{a_2a_4b}\tilde { S}_{34}^{21}(x)\Big] ,\label{gss}
 \ee
with the holomorphic functions
\begin{align} {S}_{34}^{21}(x)&~=~(1-x)^{1+i\lambda_4}x^{-1+i\lambda_2} {}_2F_1\left({2,1-i\lambda_3\atop 1+i\lambda_2};x\right)B(1-i\lambda_3,-i\lambda_4)\ ,\label{sss}\\[1mm]
\tilde{S}_{34}^{21}(x)&~=~-(1-x)^{1+i\lambda_4}x^{-1+i\lambda_2} {}_2F_1\left({2,-i\lambda_3\atop i\lambda_2};\,x\right)B(-i\lambda_3,-i\lambda_4)\ .\label{sst}\end{align}

A conformal block of a primary field with chiral weights $(h,\bar h)$ has the form \cite{Osborn:2012vt}
\be K_{34}^{21}[h,\bh]=\bx^{\bh-\bh_3-\bh_4}\hy\left({\bh-\bh_{12},\bh+\bh_{34}\atop 2\bh};\bx\right)x^{h-h_3-h_4}\hy\left({h-h_{12},h+h_{34}\atop 2h};x\right)\ ,\label{bls}\ee
where $h_{12}=h_1-h_2$ and $h_{34}=h_3-h_4$. In our case
\begin{align}\nonumber&
h_{12}=h_{1'}-h_2= 1-\textstyle \frac{i\lambda_2}{2}\ , & \bh_{12}=\bh_{1'}-\bh_2=-1-\textstyle\frac{i\lambda_2}{2}\ ,\\[1mm] &
h_{34}=h_3-h_4=\textstyle\frac{i\lambda_3}{2}-\textstyle\frac{i\lambda_4}{2}
\ , & \bh_{34}=\bh_3-\bh_4=\textstyle\frac{i\lambda_3}{2}-\textstyle\frac{i\lambda_4}{2}\ ,\\[1mm] &
h_3+h_4=2+\textstyle\frac{i\lambda_3}{2}+\textstyle\frac{i\lambda_4}{2}=2-\textstyle\frac{i\lambda_2}{2}\ , & ~~~~~~\bh_3+\bh_4=\textstyle\frac{i\lambda_3}{2}+\textstyle\frac{i\lambda_4}{2}=-\textstyle\frac{i\lambda_2}{2}\ ,\nonumber
\end{align}
where we used $\lambda_2+\lambda_3+\lambda_4=0$. It is easy to see that all conformal blocks of the correlator
 (\ref{gss}) must share common antiholomorphic weight  $\bh=1+\frac{i\lambda_2}{2}$. Indeed, its antiholomorphic part can be written as
\begin{align}
 (1-\bx)^{-1+i\lambda_4}\bx^{1+i\lambda_2}&=\bx^{1+i\lambda_2} {}_2F_1\left({2+i\lambda_2, 1-i\lambda_4\atop 2+i\lambda_2} ;\bx\right)\nonumber\\[.5mm]
&=\left.\bx^{\bar{h}-\bar{h}_3-\bar{h}_4}\hy\left({\bh-\bh_{12},\bh+\bh_{34}\atop 2\bh};\bx\right)\right|_{\bh=1+\frac{i\lambda_2}{2}}\ .\end{align}
The holomorphic parts (\ref{sss}) and (\ref{sst}) cannot be associated to a single weight, but they are so ``close'' to single blocks that they can be decomposed by using recursion relations and basic properties of hypergeometric functions.
First,
\begin{equation}\label{hy1}
\hy\left({a,b\atop c};x\right) = (1-x)^{c-a-b}\hy\left({c-a,c-b\atop c};x\right).\ee
Then, from Gau{\ss} recursion relations, it follows that
\be \hy\left({a,b\atop c-1};x\right)=\sum_{m=0}^{\infty} \frac{(a)_m (b)_m}{(c-1)_{2m}} x^m \,\hy\left({a+m,b+m\atop c+2m};x\right)\ ,\label{hy2}
\ee
where $(a)_m=\Gamma(a+n)/\Gamma(a)$ are the Pochhammer symbols, and
\be
 \hy\left({a,b\atop c};x\right)= \sum_{m=0}^{\infty}  \frac{(-1)^m (a)_m (c-b)_m}{(c)_{2m}} x^m {}_2F_1\left({a+m,b+m+1\atop c+2m+1};x\right)\label{hy3}\ee
Eqs.(\ref{hy1})-(\ref{hy3}) are completely sufficient to show that
\begin{align} {S}_{34}^{21}(x)&=\sum_{m=1}^\infty x^{h_{m}-h_3-h_4} \,a_{m} \, \hy\left(
{h_{m} -h_{12}, h_{m} + h_{34}\atop 2h_{m}}; x\right) \ ,
\label{sss1}\\[1mm]
\tilde{S}_{34}^{21}(x)&~=\sum_{m=1}^\infty x^{h_{m}-h_3-h_4} \,\tilde a_{m} \, \hy\left(
{h_{m} -h_{12}, h_{m} + h_{34}\atop 2h_{m}}; x\right) \ ,
 \label{sst1}\end{align}
where
\begin{equation}
h_{m} = m+\frac{i\lambda_2}{2}\ ,
\end{equation}
and the coefficients
\begin{align}\label{coeff1}
a_m& =\frac{m!\,\Gamma(-i\lambda_3+m)\Gamma(-i\lambda_4)}{\Gamma(i\lambda_2+2m-1)}\ ,\\[1mm]\label{coeff2}
\tilde a_m&=-a_m+(-1)^m\frac{m!\,\Gamma(-i\lambda_4+m)\Gamma(-i\lambda_3)}{\Gamma(i\lambda_2+2m-1)}\ .
\end{align}
In this way, we obtain
\be G_{34}^{21}(x,\bx)_{s,\lambda_1=0} = \sum_{m=1}^{\infty}
(a_m\, f^{a_1a_2b}f^{a_3a_4b}+\tilde a_m\, f^{a_1a_3b}f^{a_2a_4b})
K_{34}^{21}\Big[m+\frac{i\lambda_2}{2} ,1+\frac{i\lambda_2}{2}\Big].\label{gss1}
 \ee

The contributions of $t$ and $u$ channels can be analyzed in a similar way. In Eqs.(\ref{it0}), (\ref{iu0}), (\ref{tit}) and (\ref{tiu}), $t$- and $u$-channel integrals are written as functions of $1/x$ and $1-x$, respectively. In order to decompose them into  $(12\rightleftharpoons 34)_{\bm{\mathfrak{2}}}$ blocks (\ref{bls}), we need to express them as functions of $x$ and write as power series in $x$. This is easy to accomplish by using well-known hypergeometric identities. For example
\begin{align}
I_t(x) &= -\frac{(x \bar{x})^{\frac{i\lambda_2}{2}}}{x^2 \bar{x}} B(2-i\lambda_2,-i\lambda_4) \, _2F_1\left({2,2-i\lambda_2\atop 2+i\lambda_3}; \frac{1}{x}\right) \nonumber\\
&~~~=-\frac{(x \bar{x})^{\frac{i\lambda_2}{2}}}{ \bar{x}}\bigg[ B(-i\lambda_2,-i\lambda_4) \, _2F_1\left({2,1-i\lambda_3\atop 1+i\lambda_2};x\right)\\[1mm]
&~~~~\qquad~\qquad\qquad +\,(-x)^{-i\lambda_2}\frac{(1-i\lambda_2)\pi}{\sin(\pi i\lambda_2)} \, _2F_1\left({2-i\lambda_2, 1+i\lambda_4\atop 1-i\lambda_2};x\right) \bigg] .\nonumber
\end{align}
Other integrals can be transformed in a similar way. All of them contain a new class of terms with the prefactor $x^{-i\lambda_2}$. Such terms shift the spins of conformal blocks, which are always integer in the $s$ channel, by an imaginary amount $-i\lambda_2$. These new states have chiral weights $(h,\bh)= (m-\frac{i\lambda_2}{2} ,1+\frac{i\lambda_2}{2})$, therefore positive integer dimensions $\Delta=2+M$ with $M\ge 0$ and continuous complex spin $J=\Delta-2-i\lambda_2$. Are these states a part of CCFT spectrum or just a dual description of integer spins? The answer depends whether $s$, $t$ and $u$ channels of celestial amplitudes are assembled into one CCFT correlator or they are considered as distinct correlators. In the latter case, four-dimensional incoming and outgoing wave packets should be associated to different two-dimensional primary fields \cite{Strominger1910}. Then $G_{34}^{21}(x,\bx)_{s}$ encompass a full CCFT correlator and its decomposition in the ``compatible'' $(12\rightleftharpoons 34)_{\bm{\mathfrak{2}}}$ channel yields integer spin only. When celestial amplitudes are decomposed in ``incompatible'' channels, {\it e.g}.\ four-dimensional $t$ channel decomposed into $(12\rightleftharpoons 34)_{\bm{\mathfrak{2}}}$ blocks, imaginary spin states appear as a dual description of compatible channels.

The group-dependence of the correlator (\ref{gss1}) is contained in the factors
\be c_m=a_m\, f^{a_1a_2b}f^{a_3a_4b}+\tilde a_m\, f^{a_1a_3b}f^{a_2a_4b}\label{cmm} \ee
which determine the gauge group representations of  states propagating through $(12\rightleftharpoons 34)_{\bm{\mathfrak{2}}}$ conformal blocks.
While the first factor propagates the adjoint representation only, the second factor includes other representations contained in the product of two adjoint representations. For example, in the case of $SU(2)$ with $I=1$ isospin gluons,
\be f^{a_1a_3b}f^{a_2a_4b} =\delta^{a_1a_2}\delta^{a_3a_4}-\delta^{a_1a_4}\delta^{a_2a_3} \ee
therefore the blocks also include $I=0$ (trace part) and $I=2$ (symmetric traceless). Note that the factors $c_m$ (\ref{cmm}) are symmetric under $3\leftrightarrow 4$ for odd $m$ (even spin) and antisymmetric for even $m$ (odd spin), which can be seen by using Jacobi identity and Eqs.(\ref{coeff1}) and (\ref{coeff2}).

To summarize, the soft limit of $\lambda_1=0$ leads to remarkable simplifications. In this case, conformal blocks are associated to primary fields with chiral weights $(h,\bh) = (m{+}\frac{i\lambda_2}{2}\ ,1{+}\frac{i\lambda_2}{2})$, with $m\ge 1$. They have dimensions $\Delta=2+J+i\lambda_2$, where $J\ge 0$ is an integer spin. They come in all gauge group representations contained in the product of two adjoint representations.
\subsection{General case}
{}For general complex dimensions $\Delta=1+i\lambda$ (always subject to the constraint $\lambda_1+\lambda_2+\lambda_3+\lambda_4=0$), the integrals (\ref{iplus}) and (\ref{iminus}) can be expressed in terms of the Appell hypergeometric function
\be F_1\left({a;b_1,b_2\atop c}; x,y\right)=\sum_{n=0}^{\infty}\sum_{m=0}^{\infty}\frac{(a)_{n+m}(b_1)_n(b_2)_m}{(c)_{n+m}n!m!}\, x^ny^m\ .\label{appell}\ee
 Here, we focus on the $s$ channel integrals $I_s$ and $\tilde I_s$ originating from the integration region $(1,\infty)$  in $I_+$ and $\tilde I_+$, respectively. By comparing with the integral representation of $F_1$:
\be F_1\left({a;b_1,b_2\atop c}; x,y\right)=
\frac{\Gamma(c)}{\Gamma(a)\Gamma(c-a)}
\int_{0}^{1}t^{a-1}(1-t)^{c-a-1} (1-tx)^{-b_1}(1-ty)^{-b_2}\,dt\ ,\label{appell2}\ee
we find
\begin{align}	I_s(x,\bx) ~=~ & \frac{(x\bar{x})^{\frac{i\lambda_2}{2}-\frac{i\lambda_1}{2}}}{\bar{x}}
B\left(1+i \lambda_{2}+i \lambda_{4}, i \lambda_{2}+i \lambda_{3}\right) \, F_{1}\left({1+i \lambda_{2}+i \lambda_{4}; 2-i \lambda_{1},-i \lambda_{1}\atop 1-i \lambda_{1}+i \lambda_{2}} ; x, \bx\right)\ ,\\[1mm]
\tilde I_s(x,\bx) ~=~ & -\frac{(x\bar{x})^{\frac{i\lambda_2}{2}-\frac{i\lambda_1}{2}}}{\bar{x}}
B\left(i \lambda_{2}+i \lambda_{4}, i \lambda_{2}+i \lambda_{3}\right)\, F_{1}\left({i \lambda_{2}+i \lambda_{4}; 2-i \lambda_{1},-i \lambda_{1}\atop -i \lambda_{1}+i \lambda_{2}} ; x,\bx \right)\ .
\end{align}
Appell functions can be expanded into the products of hypergeometric functions by using Burchnall-Chaundy expansion \cite{BC1940}, which is exactly what we need for  conformal block decomposition:
\begin{align}\nonumber
F_1\left({a;b_1,b_2\atop c}; x,y\right) =\sum_{n=0}^\infty & {(a)_n(b_1)_n(b_2)_n(c-a)_n \over n!(c+n-1)_n (c)_{2n}}\,x^n y^n\\ &\times\hy\left({a+n,b_1+n\atop c+2n};x\right)\hy\left({a+n,b_2+n\atop c+2n};y\right)\ .
\label{F1exp}\end{align}
In our case,
\begin{align}\label{eq:I2exp}
I_s&(x,\bx) ~=~\frac{(x\bar{x})^{\frac{i\lambda_2}{2}-\frac{i\lambda_1}{2}}}{\bar{x}}
B(1+i\lambda_2+i\lambda_4,i\lambda_2+i\lambda_3)\,\nonumber\times\\[1mm] & \qquad\qquad\times \sum_{n=0}^\infty {(1+i\lambda_2+i\lambda_4)_n(2-i\lambda_1)_n(-i\lambda_1)_n(-i\lambda_1-i\lambda_4)_n \over n!(n-i\lambda_1+i\lambda_2)_n (1-i\lambda_1+i\lambda_2)_{2n}}\, \times \\[1mm]
&\times x^{n}  \hy\left({1+i\lambda_2+i\lambda_4+n,2-i\lambda_1+n \atop 1-i\lambda_1+i\lambda_2+2n};x\right)
\bx^{n} \hy\left({1+i\lambda_2+i\lambda_4+n,-i\lambda_1+n\atop 1-i\lambda_1+i\lambda_2+2n};\bx\right),\nonumber
\end{align}
and a similar expression for $\tilde I_s(x,\bx)$. From this point, conformal block decomposition proceeds as in the case of $\lambda_1=0$, by applying Eqs.(\ref{hy1})-(\ref{hy3}) to both holomorphic and antiholomorphic sides.
After a lengthy computation, we find
\begin{align} G_{34}^{21}(x,\bx)_{s} = \sum_{m,n=0}^{\infty}&
(a_{mn}\, f^{a_1a_2b}f^{a_3a_4b}+\tilde a_{mn}\, f^{a_1a_3b}f^{a_2a_4b})\nonumber\\ &\times\!
K_{34}^{21}\Big[m+1+\frac{i\lambda_2}{2}-\frac{i\lambda_1}{2} ,n+1+\frac{i\lambda_2}{2}-\frac{i\lambda_1}{2}\Big],\label{gss2}
 \end{align}
with
\begin{align}
	a_{mn}	~=~ &\frac{(2-i\lambda_1)_m(-i\lambda_1)_n\Gamma(1+m+i(\lambda_2+\lambda_4))
\Gamma(1+n+i(\lambda_2+\lambda_4)) }{\Gamma(1+2m+i(\lambda_2-\lambda_1))\Gamma(1+2n+i(\lambda_2-\lambda_1))}
	 \nonumber\\
	 &~ \quad\times\!
\sum_{r=0}^{\mathrm{min}(m,n)}\frac{(2r+i(\lambda_2-\lambda_1))\Gamma(r-i(\lambda_1+\lambda_4))  \Gamma(r+i(\lambda_2-\lambda_1))}{r! \,  \Gamma(1+r+i(\lambda_2+\lambda_4))}\ ,
\end{align}
\begin{align}
\tilde a_{mn}&	~=~ - 	\frac{(2-i\lambda_1)_m(-i\lambda_1)_n\Gamma(1+m-i(\lambda_1+\lambda_4))
\Gamma(1+n-i(\lambda_1+\lambda_4)) }{\Gamma(1+2m+i(\lambda_2-\lambda_1))
\Gamma(1+2n+i(\lambda_2-\lambda_1))}
\nonumber\\
& \times
\sum_{s=0}^{m}\sum_{t=0}^{n} (-1)^{m+n-s-t} \frac{(2s+i(\lambda_2-\lambda_1))(2t+i(\lambda_2-\lambda_1))
\Gamma(s+i(\lambda_2+\lambda_4))\Gamma(t+i(\lambda_2+\lambda_4)) }{\Gamma(1+s-i(\lambda_1+\lambda_4))
\Gamma(1+t-i(\lambda_1+\lambda_4))} \nonumber\\[.5mm]
&~ \quad\qquad \times\!
\sum_{r=0}^{\mathrm{min}(s,t)}\frac{(2r-1+i(\lambda_2-\lambda_1))\Gamma(r-i(\lambda_1+\lambda_4)) \Gamma(r-1+i(\lambda_2-\lambda_1))}{r! \,   \Gamma(r+i(\lambda_2+\lambda_4))} \ .
\end{align}
The above expressions can be simplified to
\begin{align}
	a_{mn}
~=~ &\frac{(2-i\lambda_1)_m(-i\lambda_1)_n\Gamma(1+m+i(\lambda_2+\lambda_4))
\Gamma(1+n+i(\lambda_2+\lambda_4)) }{\Gamma(1+2m+i(\lambda_2-\lambda_1))\Gamma(1+2n+i(\lambda_2-\lambda_1))} \nonumber\\
&~\quad\times \frac{\Gamma(1+N+i\lambda_2-i\lambda_1)
\Gamma(1+N-i\lambda_1-i\lambda_4)}{N!(-i\lambda_1-i\lambda_4)
\Gamma(1+N+i\lambda_2+i\lambda_4)},
\end{align}
where $N={\rm min}(m,n)$, and
\be
\tilde a_{mn} ~=~  -a_{mn}
+(-1)^{m-n+1} a_{mn}(3\leftrightarrow 4) \ ,
\ee
where $a_{mn}(3\leftrightarrow 4)$ is obtained by exchanging 3 and 4 in the expression for $a_{mn}$. Note that in the limit of $\lambda_1=0$, $a_{mn}=0$ for $n>0$ because then $(-i\lambda_1=0)_n=0$. Then also $N=0$ and the result agrees with Eqs.(\ref{coeff1}) and (\ref{coeff2}).

To summarize, we find that the conformal blocks of celestial gluon amplitudes describe primary fields with chiral weights $(h,\bh) = (m{+}\frac{i\lambda_2}{2}{-}\frac{i\lambda_1}{2}\ ,n{+}\frac{i\lambda_2}{2}{-}\frac{i\lambda_1}{2})$, with integers $m,n\ge 1$. They have dimensions $\Delta=2+M+i(\lambda_2-\lambda_1)$ where $M\ge 0$ is integer, and spin $J=-M, -M{+}2,\dots,M{-}2, M$. They come in all gauge group representations contained in the product of two adjoint representations.

Conformal block decomposition of $G_{34}^{21}(x,\bx)_{s}$ in incompatible channels is more complicated. The correlator does not factorize into holomorphic and antiholomorphic parts in any simple way. Unlike in the $\lambda_1=0$ limit, there is unbounded spectrum of complex spin for each conformal dimension.

\section{Conclusions}
In this work, we exhibited conformal blocks of four-gluon amplitudes with one gluon replaced by a shadow field. Since a shadow transform of a shadow field gives back the same field, the celestial amplitude with four gluon fields can be recovered by applying subsequent shadow transformation to the blocks.

{}Four-dimensional crossing symmetry connects the amplitudes describing scattering processes in distinct physical channels. In four-gluon celestial amplitudes, the intervals of the cross ratio $z>1$, $0<z<1$ and $z<0$ describe  processes with $s>0$, $t>0$ and $u>0$, respectively.
When one such amplitude is decomposed into conformal blocks in a compatible channel,
for instance $(12 \leftrightharpoons 34)_{\bm{\mathfrak{4}}}$ decomposed into  $(12 \leftrightharpoons 34)_{\bm{\mathfrak{2}}}$
blocks, only integer spin states appear in the spectrum. We discovered primary fields with dimensions $\Delta=2+M+i\lambda$, where $M\ge 0$ is an integer, and spin $J=-M,-M+2,\dots,M-2,M$.
The states with  complex spin, but with positive integer dimensions,  appear in incompatible channels only, in a dual channel description of integer spin states.

What is the origin of an infinite tower of primary fields in CCFT? What is their four-dimensional interpretation? Some of them are certainly the ``supertranslation modes'' of gluon fields. While supertranslations shift conformal field dimensions \cite{Stieberger1812}, they do not change spin or gauge group representations. The presence of higher spin fields in various group representations indicates that CCFT symmetries go far beyond the BMS symmetry. A related question is what is the role of four-dimensional conformal symmetry  enjoyed by Yang-Mills theory at the tree level? How is it  realized at the level of these higher spin states? A detailed analysis of conformal blocks should help answering all these questions.
\vskip 1mm
\noindent {\bf Acknowledgments}\\[2mm]
We are grateful to V.\ Dotsenko, H.\ Osborn, D.\ Stanford,  B.\ Van Rees, and M.\ Zlotnikov  for useful correspondence.
This material is based in part upon work supported by the National Science Foundation
under Grant Number PHY--1913328.
Any opinions, findings, and conclusions or recommendations
expressed in this material are those of the authors and do not necessarily
reflect the views of the National Science Foundation.


\newpage




\begin{thebibliography}{99}
\bibitem{Bondi:1962px}
  H.~Bondi, M.~G.~J.~van der Burg and A.~W.~K.~Metzner,
  ``Gravitational waves in general relativity. 7. Waves from axisymmetric isolated systems,''
  Proc.\ Roy.\ Soc.\ Lond.\ A {\bf 269}, 21 (1962).
\bibitem{Sachs:1962wk}
  R.~K.~Sachs,
  ``Gravitational waves in general relativity. 8. Waves in asymptotically flat space-times,''
  Proc.\ Roy.\ Soc.\ Lond.\ A {\bf 270}, 103 (1962).
\bibitem{Barnich:2009se}
  G.~Barnich and C.~Troessaert,
  ``Symmetries of asymptotically flat 4 dimensional spacetimes at null infinity revisited,''
  Phys.\ Rev.\ Lett.\  {\bf 105}, 111103 (2010)
  %doi:10.1103/PhysRevLett.105.111103
  [arXiv:0909.2617 [gr-qc]].
\bibitem{Strominger:2013jfa}
  A.~Strominger,
  ``On BMS Invariance of Gravitational Scattering,''
  JHEP {\bf 1407}, 152 (2014)
  %doi:10.1007/JHEP07(2014)152
  [arXiv:1312.2229 [hep-th]].
\bibitem{Strominger:2017zoo}
  A.~Strominger,
  {\it Lectures on the Infrared Structure of Gravity and Gauge Theory},
Princeton University Press (2018)
  [arXiv:1703.05448 [hep-th]].
\bibitem{Strominger1706}
S.~Pasterski, S.~H.~Shao and A.~Strominger,
``Gluon Amplitudes as 2d Conformal Correlators,''
Phys. Rev. D \textbf{96} (2017) no.8, 085006
[arXiv:1706.03917 [hep-th]].
\bibitem{Pasterski1705}
S.~Pasterski and S.~H.~Shao,
``Conformal basis for flat space amplitudes,''
Phys. Rev. D \textbf{96} (2017) no.6, 065022
[arXiv:1705.01027 [hep-th]].
\bibitem{Weinb}
S.~Weinberg,
``Infrared photons and gravitons,''
Phys. Rev. \textbf{140} (1965), B516-B524
\bibitem{Cachazo:2014fwa}
F.~Cachazo and A.~Strominger,
``Evidence for a New Soft Graviton Theorem,''
[arXiv:1404.4091 [hep-th]].
%297 citations counted in INSPIRE as of 05 Mar 2021
%\bibitem{He:2014laa}
\bibitem{Strominger1401}
T.~He, V.~Lysov, P.~Mitra and A.~Strominger,
``BMS supertranslations and Weinberg's soft graviton theorem,''
JHEP \textbf{05} (2015), 151
[arXiv:1401.7026 [hep-th]].
%344 citations counted in INSPIRE as of 28 Jun 2020

%\cite{Kapec:2016jld}
%\bibitem{Kapec:2016jld}
\bibitem{Strominger1609}
D.~Kapec, P.~Mitra, A.~M.~Raclariu and A.~Strominger,
``2D Stress Tensor for 4D Gravity,''
Phys. Rev. Lett. \textbf{119} (2017) no.12, 121601
[arXiv:1609.00282 [hep-th]].
%71 citations counted in INSPIRE as of 28 Jun 2020




\bibitem{Strominger1810}
L.~Donnay, A.~Puhm and A.~Strominger,
``Conformally Soft Photons and Gravitons,''
JHEP \textbf{01} (2019), 184
[arXiv:1810.05219 [hep-th]].
%31 citations counted in INSPIRE as of 28 Jun 2020








\bibitem{Fan1903}
W.~Fan, A.~Fotopoulos and T.~R.~Taylor,
``Soft Limits of Yang-Mills Amplitudes and Conformal Correlators,''
JHEP \textbf{05} (2019), 121
[arXiv:1903.01676 [hep-th]].
%17 citations counted in INSPIRE as of 28 Jun 2020



\bibitem{Strominger1904}
M.~Pate, A.~M.~Raclariu and A.~Strominger,
``Conformally Soft Theorem in Gauge Theory,''
Phys. Rev. D \textbf{100} (2019) no.8, 085017
[arXiv:1904.10831 [hep-th]].


%\cite{Adamo:2019ipt}
%\bibitem{Adamo:2019ipt}
\bibitem{Mason1905}
T.~Adamo, L.~Mason and A.~Sharma,
``Celestial amplitudes and conformal soft theorems,''
Class. Quant. Grav. \textbf{36} (2019) no.20, 205018
[arXiv:1905.09224 [hep-th]].
%17 citations counted in INSPIRE as of 28 Jun 2020



%\cite{Puhm:2019zbl}
%\bibitem{Puhm:2019zbl}
\bibitem{Puhm1905}
A.~Puhm,
``Conformally Soft Theorem in Gravity,''
JHEP \textbf{09}, 130 (2020)
[arXiv:1905.09799 [hep-th]].


%\cite{Guevara:2019ypd}
%\bibitem{Guevara:2019ypd}
\bibitem{Guevara1906}
A.~Guevara,
``Notes on Conformal Soft Theorems and Recursion Relations in Gravity,''
[arXiv:1906.07810 [hep-th]].
%13 citations counted in INSPIRE as of 28 Jun 2020
\bibitem{Foto1906}
A.~Fotopoulos and T.~R.~Taylor,
``Primary Fields in Celestial CFT,''
JHEP \textbf{10} (2019), 167
[arXiv:1906.10149 [hep-th]].

%\cite{Pasterski:2017kqt}
%\bibitem{Pasterski:2017kqt}
\bibitem{Foto1912}
A.~Fotopoulos, S.~Stieberger, T.~R.~Taylor and B.~Zhu,
``Extended BMS Algebra of Celestial CFT,''
JHEP \textbf{03}, 130 (2020)
[arXiv:1912.10973 [hep-th]].
\bibitem{tt}  T.R.~Taylor,
  ``A Course in Amplitudes,''
Phys.\ Rept.\  {\bf 691}, 1 (2017).
[arXiv:1703.05670 [hep-th]].
\bibitem{Strominger1910}
M.~Pate, A.~M.~Raclariu, A.~Strominger and E.~Y.~Yuan,
``Celestial Operator Products of Gluons and Gravitons,''
[arXiv:1910.07424 [hep-th]].
%\cite{Ebert:2020nqf}%\cite{Banerjee:2020kaa}
\bibitem{Ebert:2020nqf}
S.~Ebert, A.~Sharma and D.~Wang,
``Descendants in celestial CFT and emergent multi-collinear factorization,''
%``Descendants in celestial CFT and emergent multi-collinear factorization,''
JHEP \textbf{03}, 030 (2021)
[arXiv:2009.07881 [hep-th]].
%2 citations counted in INSPIRE as of 28 Feb 2021


%\cite{Banerjee:2020kaa}
\bibitem{Banerjee:2020kaa}
S.~Banerjee, S.~Ghosh and R.~Gonzo,
``BMS symmetry of celestial OPE,''
%``BMS symmetry of celestial OPE,''
JHEP \textbf{04}, 130 (2020)
[arXiv:2002.00975 [hep-th]].
\bibitem{DiF} P. Di Francesco, P. Mathieu, D. S\'en\'echal, ``Conformal Field Theory,'' Springer (1997).

\bibitem{Lam:2017ofc}
H.~T.~Lam and S.~H.~Shao,
``Conformal Basis, Optical Theorem, and the Bulk Point Singularity,''
Phys. Rev. D \textbf{98} (2018) no.2, 025020
[arXiv:1711.06138 [hep-th]].
%27 citations counted in INSPIRE as of 28 Feb 2021


%\bibitem{Nandan:2019jas}
\bibitem{Volovich1904}
D.~Nandan, A.~Schreiber, A.~Volovich and M.~Zlotnikov,
``Celestial Amplitudes: Conformal Partial Waves and Soft Limits,''
JHEP \textbf{10} (2019), 018
[arXiv:1904.10940 [hep-th]].
%17 citations counted in INSPIRE as of 28 Jun 2020


%\cite{Law:2020xcf}
\bibitem{Law:2020xcf}
Y.~T.~A.~Law and M.~Zlotnikov,
``Relativistic partial waves for celestial amplitudes,''
JHEP \textbf{11} (2020), 149
[arXiv:2008.02331 [hep-th]].
\bibitem{Osborn:2012vt}
H.~Osborn,
``Conformal Blocks for Arbitrary Spins in Two Dimensions,''
Phys. Lett. B \textbf{718}, 169-172 (2012)
[arXiv:1205.1941 [hep-th]].
%\cite{Fotopoulos:2020bqj}
\bibitem{Fotopoulos:2020bqj}
A.~Fotopoulos, S.~Stieberger, T.~R.~Taylor and B.~Zhu,
``Extended Super BMS Algebra of Celestial CFT,''
JHEP \textbf{09}, 198 (2020)
[arXiv:2007.03785 [hep-th]].
%\cite{Sachs:1962wk}
%\cite{Bondi:1962px}

\bibitem{Law:2019glh}
Y.~T.~A.~Law and M.~Zlotnikov, ``Poincar\'e constraints on celestial amplitudes,''
JHEP \textbf{03}, 085 (2020)
[erratum: JHEP \textbf{04}, 202 (2020)]
[arXiv:1910.04356 [hep-th]].
\bibitem{spas} S. Pasterski, {https://physicsgirl.com/ss.pdf}.
\bibitem{DelDuca:1999rs}
V.~Del Duca, L.~J.~Dixon and F.~Maltoni,
``New color decompositions for gauge amplitudes at tree and loop level,''
Nucl. Phys. B \textbf{571}, 51-70 (2000)
[arXiv:hep-ph/9910563 [hep-ph]].
\bibitem{Stieberger1812}
S.~Stieberger and T.~R.~Taylor,
``Symmetries of Celestial Amplitudes,''
Phys. Lett. B \textbf{793} (2019), 141-143
[arXiv:1812.01080 [hep-th]].
\bibitem{BC1940}
 J.L.\ Burchnall and  T.W.\ Chaundy, ``Expansions of Appell's Double Hypergeometric Functions,'' Quart. J. Math. (Oxford),  vol.\ \textbf{11}, pp. 249-270 (1940);
`` Expansions of Appell's Double Hypergeometric Functions (II),'' Quart. J. Math. (Oxford), vol.\ \textbf{12}, pp. 112-128 (1941).


\end{thebibliography}
\end{document}